# Structure of strongly interacting polyelectrolyte diblock copolymer micelles

A.V. Korobko and W. Jesse
*Leiden Institute of Chemistry, Leiden University, 2300 RA Leiden, the Netherlands*

A. Lapp
*Laboratoire Léon Brillouin, CEA/CNRS, 91191 Gif-sur-Yvette Cedex, France*

S.U. Egelhaaf
*School of Physics and School of Chemistry, The University of Edinburgh, Edinburgh EH9 3JZ, United Kingdom*

J.R.C. van der Maarel[a]
*Leiden Institute of Chemistry, Leiden University, 2300 RA Leiden, the Netherlands and Department of Physics, National University of Singapore, Singapore 117542*

(20 September 2004)

The structure of spherical micelles of the diblock poly(styrene-*block*-acrylic acid) [PS-*b*-PA] copolymer in water was investigated up to concentrations where the polyelectrolyte coronal layers have to shrink and/or interpenetrate in order to accommodate the micelles in the increasingly crowded volume. We obtained the partial structure factors pertaining to the core and corona density correlations with small angle neutron scattering (SANS) and contrast matching in the water. The counterion structure factor was obtained with small angle X-ray scattering (SAXS) with a synchrotron radiation source. Furthermore, we have measured the flow curves and dynamic visco-elastic moduli. The functionality of the micelles is fixed with a 9 nm diameter PS core and a corona formed by around 100 PA arms. As shown by the SAXS intensities, the counterions are distributed in the coronal layer with the same density profile as the corona forming segments. Irrespective ionic strength and micelle charge, the corona shrinks with increasing packing fraction. At high charge and minimal screening conditions, the polyelectrolyte chains remain almost fully stretched and they interdigitate once the volume fraction exceeds the critical value 0.53±0.02. Interpenetration of the polyelectrolyte brushes also controls the fluid rheology: the viscosity increases by 3 orders of magnitude and the parallel frequency scaling behavior of the dynamic moduli suggests the formation of a physical gel. In excess salt, the coronal layers are less extended and they do not interpenetrate in the present concentration range.

[a] Author for correspondence. Voice: +31 715274543, Fax: +31 71 5274603, E-mail: j.maarel@chem.leidenuniv.nl

## I. INTRODUCTION

Polyelectrolyte diblock copolymers have found widespread applications from the stabilization of colloidal suspensions, through encapsulation and delivery of bioactive agents, to the control of gelation, lubrication, and flow behavior.[1,2] In water or aqueous solution, the hydrophobic attachment provides a mechanism for self-assembling and mesoscopic structures are formed. These structures can be classified according their morphologies, including spherical and cylindrical micelles, as well as lamellar and vesicular structures.[3,4] The polyelectrolyte chains are anchored at the hydrophobic micro-domains and they form an interfacial brush. The key concept in understanding of the functioning of this class of materials is the structure of the polyelectrolyte brush in terms of the polymer and counterion density profiles.[5] In contrast to neutral brushes, stretching of the polyelectrolyte brush is primarily affected by the osmotic pressure exerted by the counterions adsorbed in the layer, rather than the repulsion between monomers.[6,7,8]

Spherical micelles of polyelectrolyte diblock copolymers typically consist of a neutral core formed by the self-assembled neutral blocks and surrounded by a polyelectrolyte coronal layer. In particular, when the functionality (*i.e.*, aggregation number) is fixed due to the high glass temperature of the core, these micelles provide an excellent model system to investigate the properties of the corona without complications related to copolymer rearrangements. For individual micelles, the corona size and its relation to charge, screening, and counterion distribution have been investigated with scattering techniques. The main results are osmotic star-branched polyelectrolyte behavior, full corona chain stretching at high charge and minimal screening conditions, and robustness of the coronal layer against the salinity generated by the addition of salt.[9,10,11] At low degrees of ionization, the corona charges migrate to the outer micelle region due to the recombination/dissociation balance of weak polyacid (charge annealing).[12] It was also found that the counterion radial density profile is very close to the one for the corona forming copolymer segments and that most, if not all counterions are adsorbed in the coronal layer.[13,14]

Despite the considerable body of experimental work, not much is known about the organization among micelles and how the coronal layers respond when the micelles interact. Electrostatic interactions are anticipated to be of minor importance, due to the almost complete neutralization of the micelles by trapping of the counterions in the polyelectrolyte brush. In particular, the extent to which the relatively dense brushes of copolymer micelles contract and/or interdigitate is an open question.[15] Many of the technological applications are derived from the interaction between polyelectrolyte brushes in concentrated systems and it is our contention that interdigitation is important in explaining the fluid behavior. In a recent letter, we reported small angle neutron scattering (SANS) experiments on a model system of spherical micelles up to concentrations where the coronas have to shrink and/or interpenetrate in order to accommodate the micelles in the increasingly crowded volume.[16] It was observed that, irrespective of ionic strength, the corona shrinks with increasing packing fraction. Furthermore, at high charge and minimal screening conditions, the corona layers interpenetrate once the volume fraction exceeds a certain critical value. In this paper, we give a more detailed account of the data analysis and report more results, including small angle X-ray (SAXS) experiments for the determination of the counterion distribution as well as flow measurements to characterize fluid rheology.

We studied micelles formed by poly(styrene-*block*-acrylic acid) [PS-*b*-PA] with degrees of polymerization 20 and 85 of the PS and PA blocks, respectively. At ambient temperature, the PS core is in a glassy state, which results in micelles with fixed core size and functionality (the glass temperature of PS, $T_g$ = 363 K). The PA corona charge is pH dependent and can be varied between almost zero and full (100%) charge where every monomer carries an ionized group. In the scattering experiments, we focus on the counterion distribution in the coronal layer, the contraction of the corona before overlap, interdigitation at high packing fraction, and the relation with charge and electrostatic screening.[17,18] The core and corona structure factors, as obtained from SANS and contrast matching in water, are interpreted in terms of core size, statistical properties of the corona-forming segments, and the thickness of the coronal layer. The ion distribution in the coronal layer is gauged from a comparison of the corona structure factor and the SAXS intensity dominated by the scattering of the relatively heavy Cs$^+$ counterions. Comparison of the micelle diameter from the form factor analysis with the effective diameter from the micellar center of mass structure factor will then show the extent to which the coronal layers interpenetrate. To further investigate interdigitation and the possible formation of an interconnected network of micelles, we have measured the shear rate dependence of the viscosity and frequency sweeps of the visco-elastic storage and loss moduli of some representative samples.

## II. SCATTERING ANALYSIS

### A. From intensities to structure factors

The structure factors describing the density correlations of the PS and PA copolymer blocks are obtained from SANS. For a diblock PS($N_{PS}$)-b-PA($N_{PA}$) copolymer solution, with $N_{PS}$ and $N_{PA}$ the number of monomers of the PS and PA block, respectively, it is convenient to consider the blocks as the elementary scattering units.[19] Every PS block is attached to a PA block, and, hence, the macroscopic block concentrations exactly match the copolymer concentration $\rho_{PS} = \rho_{PA} = \rho$. The distribution of the counterions along the micelle radius equals the one for the PA monomers, as shown by previous SANS work of samples with isotopically labeled counterions and the SAXS experiments described below.[13] We will consider, accordingly, the diblock copolymer solution as an effective 3-component system, *i.e.*, the core PS block, the corona PA block with neutralizing counterions, and the solvent. The coherent part of the solvent corrected scattered intensity is given by the sum of 3 partial structure factors describing the density correlations among the PS and PA blocks:

$$I(q)/\rho = \bar{b}_{PS}^2 N_{PS}^2 S_{PS}(q) + 2\bar{b}_{PS}\bar{b}_{PA} N_{PS} N_{PA} S_{PS-PA}(q) + \bar{b}_{PA}^2 N_{PA}^2 S_{PA}(q) \quad (1)$$

with the block *monomer* scattering length contrasts $\bar{b}_{PS}$ and $\bar{b}_{PA}$, respectively. The scattering length density of the PA block $\bar{b}_{PA}$ is calculated by taking the relevant average of the values pertaining to PA in its acid and neutralized forms, respectively. Momentum transfer $q$ is defined

*Polyelectrolyte copolymer micelles*

by the wavelength $\lambda$ and scattering angle $\theta$ between the incident and scattered beam according to $q = 4\pi/\lambda \sin(\theta/2)$. The partial structure factors $S_{ij}(q)$ are the spatial Fourier transforms of the *block* density correlation functions

$$S_{ij}(q) = \rho^{-1} \int_V d\vec{r} \exp(-i\vec{q} \cdot \vec{r}) \langle \rho_i(0) \rho_j(\vec{r}) \rangle \tag{2}$$

with $i, j$ = PS, PA ($S_{ii}$ is abbreviated as $S_i$). In an $H_2O/D_2O$ solvent mixture, the SANS scattering length contrast is given by

$$\bar{b}_i = b_i - b_s \bar{v}_i / \bar{v}_s \;,\; b_s = X(D_2O) b_{D_2O} + (1 - X(D_2O)) b_{H_2O} \tag{3}$$

with $X(D_2O)$ the $D_2O$ mole fraction. The monomer ($i$) and solvent ($s$) have scattering lengths $b_i$ and $b_s$ and partial molar volumes $\bar{v}_i$ and $\bar{v}_s$, respectively. In our SANS experiments, the core and corona structure factors are obtained from the intensities by contrast variation in the water, *i.e.*, by adjusting the solvent scattering lengths $b_s$.

In a selective solvent, the copolymers form spherical aggregates with a hydrophobic PS block core and a polyelectrolyte PA block corona. If the radial density of the corona is assumed to be invariant to fluctuations in inter-micelle separation, the structure factor Eq. (2) takes the form

$$S_{ij}(q) = N_{ag}^{-1} F_i(q) F_j(q) \; S_{cm}(q) \tag{4}$$

with the micelle aggregation number $N_{ag}$, the form factor amplitude $F_i(q)$, and the micelle center of mass structure factor $S_{cm}(q)$. In the absence of interactions between the micelles and/or at sufficiently high values of momentum transfer $S_{cm}(q)$ reduces to unity. The form factor amplitude $F_i(q)$ can be expressed in terms of the *radial* core ($i$ = PS) or corona ($i$ = PA) density $\rho_i(r)$

$$F_i(q) = \int_{V_{micelle}} d\vec{r} \exp(-i\vec{q} \cdot \vec{r}) \rho_i(\vec{r}) = \int dr \sin(qr)/(qr) 4\pi r^2 \rho_i(r) \tag{5}$$

The scattering amplitudes are normalized to $N_{ag}$ at $q = 0$.

The factorization of the structure factors into the intra-micelle form factor amplitudes $F_i(q)$ and the inter-micelle center of mass structure factor $S_{cm}(q)$ according to Eq. (4) is important in recognizing certain relations between the different partial structure factors and the data analysis procedure. The center of mass structure factor $S_{cm}(q)$ is positive definite, since it represents a

*Polyelectrolyte copolymer micelles*

scattered intensity (*i.e.*, a squared amplitude). The intensities Eq. (1) can now be expressed in terms of 2 factors $u_i(q)$ rather than 3 partial structure factors $S_{ij}(q)$ ($i, j = PS, PA$):

$$I(q)/\rho = \left[\bar{b}_{PS} N_{PS} u_{PS}(q) + \bar{b}_{PA} N_{PA} u_{PA}(q)\right]^2 \ , \ u_i(q) = \left[S_{cm}(q)/N_{ag}\right]^{1/2} F_i(q) \quad (6)$$

As shown in previous work, explicit use of Eq. (4) in the data analysis procedure according to Eq. (6) is consistent with a 3-parameter fit of *all* partial structure factors.[11,12,13] Furthermore, the concomitant reduction in number of adjustable parameters results in improved statistical accuracy in the derived structure factors.

In the case of our SAXS experiments, we have neutralized the polyelectrolyte copolymer with CsOH. Since the Cs$^+$ ion is much heavier (atomic number $Z = 55$) than the organic copolymer atoms, the scattering is dominated by the counterions in the coronal layer. Accordingly, the SAXS intensity is directly proportional to the counterion structure factor, but there is also a small contribution from the copolymer (see below).

**B. Solution structure factor**

For polyelectrolyte copolymer micelles, an analytic expression for the center of mass solution structure factor is not available. We have analyzed the data with a hard sphere potential and the Percus-Yevick approximation for the closure relation.[20] The solution structure factor has the form

$$S_{cm}^{-1}(q) - 1 = 24\phi\left[\alpha f_1(D_{hs}q) + \beta f_2(D_{hs}q) + \phi\alpha f_3(D_{hs}q)/2\right] \quad (7)$$

with

$$\alpha = \frac{(1+2\phi)^2}{(1-\phi)^4} \ , \ \beta = -\frac{3\phi(2+\phi)^2}{2(1-\phi)^4} \quad (8)$$

and

$$\begin{aligned} f_1(x) &= (\sin(x) - x\cos(x))/x^3 \\ f_2(x) &= (2x\sin(x) - (x^2 - 2)\cos(x) - 2)/x^4 \\ f_3(x) &= ((4x^3 - 24x)\sin(x) - (x^4 - 12x^2 + 24)\cos(x) + 24)/x^6 \end{aligned} \quad (9)$$

The fit parameters are the hard sphere diameter $D_{hs}$ and the volume fraction $\phi = \pi/6 D_{hs}^3 \rho_{mic}$ with micelle density $\rho_{mic}$. The hard sphere diameter should be interpreted as an effective diameter; its value could be smaller than the outer micelle diameter if interpenetration occurs. Furthermore, it is known that for soft objects the hard sphere potential does not correctly predict the relative amplitudes of the primary and higher order correlation peaks.[21] As will be discussed below, we have also tested a sticky hard sphere model and a repulsive screened Coulomb potential.[22,23] However, the effect of electrostatic interaction among the micelles was found to be modest, which is attributed to the fact that almost all neutralizing counterions are confined in the coronal layer.[13,14]

## C. Core and corona form factors

The core can be described by a homogeneous dense sphere with density $\rho_{ps}$ and diameter $D_{core}$. Accordingly, the radial PS block density is uniform for $0 \leq 2r \leq D_{core}$ and given by $\rho_{PS}(r)\pi D_{core}^3/6 = N_{ag}$ and zero for $2r > D_{core}$. For such uniform profile, the core scattering amplitude reads

$$F_{PS}(q) = N_{ag} 3\left(\sin(qD_{core}/2) - (qD_{core}/2)\cos(qD_{core}/2)\right)/(qD_{core}/2)^3 \qquad (10)$$

Expressions for the scattering amplitude of Gaussian chains with constant density in the coronal layer (and the interference with the spherical core) are available in the literature.[24] However, due to the relatively small core size and the mutual segment repulsion induced by the charge, the density in the coronal layer is non-uniform and varies along with the radius away from the core. To describe the corona structure we will adopt an algebraic radial PA block density profile

$$\rho_{PA}(r) = \rho_{PA}^0 \left(2r/D_{core}\right)^{-\alpha}, \quad D_{core} < 2r < D_{mic} \qquad (11)$$

where corona chain statistics determines the value of $\alpha$ and $\rho_{PA}^0$ is the density at the core - corona interface. The latter interfacial density is related to the outer micelle diameter $D_{mic}$ through the normalization requirement (*i.e.*, by integration of the radial profile)

$$\pi\left(D_{mic}^{3-\alpha} D_{core}^{\alpha} - D_{core}^3\right) = 2(3-\alpha) N_{ag} \qquad (12)$$

We will calculate the corona form factor amplitude with algebraic profile Eq. (11) by numeric integration, although analytical expressions are available.[25] The core PS and corona PA *form* factors are related to the square of the scattering amplitudes and take the form

$$P_{PS}(q) = F_{PS}^2(q)/N_{ag}, \quad P_{PA}(q) = F_{PA}^2(q)/N_{ag} \tag{13}$$

The algebraic profile Eq. (11) accounts for the *average* corona density scaling and neglects corona chain *fluctuations*. The effect of fluctuations on the scattering behavior is important when the momentum transfer is on the order of the intermolecular correlation distance within the corona. Furthermore, they contribute to the corona structure factor (=$S_{PA}$) only, the cross term $S_{PS-PA}$ is unaffected due to the heterodyne interference between the amplitudes scattered by the homogeneous core and heterogeneous corona.[26,27]

## III. CORONA CHAIN STATISTICS

The value of the scaling exponent $\alpha$ in Eq. (11) is determined by the chain statistics in the coronal layer. In the present contribution, the corona statistics is gauged from the scaling approaches for star-branched polyelectrolytes.[6,7,8] These polymers can also serve as a model for spherical diblock copolymer micelles; the presence of the core does not invalidate the scaling results. The fact that the coronal region cannot extend right to the center of the micelle merely sets a certain minimum correlation length (*i.e.*, blob size) at the core−corona interface.

It is convenient to start the analysis of the corona statistics from micelles with a large fraction of ionized groups and no added salt. In this situation, charge-annealing effects are unimportant, most of the counterions are trapped in the coronal layer, and the concomitant osmotic pressure gives the main contribution to the corona stretching force. The radial scaling of the correlation length can be derived from the balance of the elastic, conformational, stretching force and the osmotic pressure exerted by the counterions. Since the fraction of trapped counterions does not vary along the radius, the correlation length is constant. The formation of radial strings of blobs of uniform size and, hence, uniform mass per unit length results in an outer−coronal density scaling exponent $\alpha = 2$. Due to space restrictions, in the inner−coronal region the correlation length is expected to decrease (with $\alpha$ on the order of unity). However, in previous work we found no evidence for this effect, because for sufficiently high charge fraction the correlation length is smaller than the average distance between the chains at the core−corona interface as set by the grafting density.[12] Our fully and 50% ionized samples are in the osmotic regime, and, without added salt, the chains in the coronal layer are near 100% stretched with a density scaling proportional to the inverse second power of the radius away from the core ($\alpha = 2$).

An additional screening of Coulomb interaction becomes important when the concentration of added salt exceeds the concentration of counterions in the coronal layer. The corona stretching force is now proportional to the difference in osmotic pressure of co- and counterions inside and outside the micelle. This difference in osmotic pressure can be obtained by employing the local electroneutrality condition and Donnan salt partitioning between the micelle and the bulk of the solution.[8] An increase in salt concentration results in a gradual contraction of the micelle according to $D_{mic} \sim C_s^{-1/5}$, because of additional screening of the Coulomb repulsion among the ionized polyelectrolyte block monomers (*i.e.*, a decrease in electrostatic excluded volume interactions). In the salt dominated regime, the radial decay of the monomer density is described by the same exponent $\alpha = 4/3$ as in neutral star-branched polymers with short-range excluded volume interactions in a good solvent.[28] However, in contrast to neutral stars, the elastic blobs in screened polyelectrolyte micelles have a blob-size scaling exponent 2/3 rather than unity and, hence, they are not closely packed.[8]

PA is a weak polyacid and at low degree of neutralization, the effects of charge annealing are important. Because of the dissociation and recombination balance, the charge fraction and, hence, the local tension in the branches increase with increasing distance away from the core. As the branches become more extended with increasing *r*, the correlation length *decreases* and the monomer density decays faster. The scaling exponent $\alpha$ takes the value 8/3 or 5/2 without or with volume interactions, respectively.[8] Although the addition of salt might shift the recombination–dissociation balance, the corona scaling behavior is unaffected. The additional screening results, however, in a contraction of the coronal layer according to $D_{mic} \sim C_s^{-1/5}$, as in the case of highly charged micelles.

## IV. EXPERIMENTAL SECTION

### A. Chemicals and solutions

PS-*b*-NaPA was purchased from Polymer Source Inc., Dorval, Canada. The number average degrees of polymerization of the PS and PA blocks are 20 and 85, respectively. PS-*b*-NaPA was brought in the acid form by dissolving it in 0.1 M HCl and extensive dialysis against water (purified by a Millipore system with conductivity less than $1\times10^{-6}$ $\Omega^{-1}cm^{-1}$). The residual sodium content in PS-*b*-PAA was checked by atomic absorption spectroscopy and was less than 0.001. Solutions were prepared by dissolving freeze-dried PS-*b*-PAA in pure water and/or $D_2O$ at 350 K under continuous stirring for 6 hours. Furthermore, to break up clusters of micelles, the solutions were sonicated (Bransonic 5200) for 30 minutes at room temperature. The sonication power was relatively low (190 W) and, hence, without risk of damage or decomposition of the block copolymers. Copolymer concentrations were determined by potentiometric titration with NaOH (Titrisol, Merck). The solutions were subsequently neutralized with NaOH or CsOH to a degree of neutralization *DN*. The degree of neutralization is the molar ratio of (added) alkali and polyacid monomer.

### B. Neutron scattering

For neutron scattering, 6 sets of solutions with 100, 50, and 10% corona charge (degree of neutralization $DN$ = 1.0, 0.5, and 0.1, respectively) were prepared: 3 sets without added salt, in another 3 the salt concentration is 1.0 M (100 and 50% charge) or 0.04 M (10% charge). We have used KBr instead of NaCl to minimize an incoherent scattering contribution related to the salt. Each set was prepared with 4 copolymer concentrations ranging from the dilute to the dense regime, where the coronas should interpenetrate if they do not shrink. Furthermore, we applied contrast variation with 4 solvent compositions. For this purpose, all solutions were prepared in $H_2O$ and $D_2O$ and subsequently mixed by weight to obtain 4 different $H_2O/D_2O$ solvent compositions. The solvent compositions were checked by the values for transmission. Scattering length contrasts were calculated with Eq. (3) and the parameters in Table I and are collected in Table II. For each degree of neutralization, the corona scattering length contrast has been calculated by taking the relevant average of the PAA and NaPA contrast parameters, $\bar{b}_{PAA}$ and $\bar{b}_{NaPA}$, respectively. Reference solvent samples with matching $H_2O/D_2O$ composition were also prepared. Standard quartz sample containers with 0.1 cm (for samples in pure $H_2O$) or 0.2 cm path length were used.

SANS was measured with the D22 and PAXY diffractometers situated on the cold sources of the Institute Max von Laue – Paul Langevin (Grenoble) and Laboratoire Léon Brillouin (CE de Saclay), respectively. The temperature was kept at 293 K. The sample sets with 100 % charged micelles without added salt as well as the 10% charged micelles in 0.04 M KBr were measured with the D22 instrument in 2 different configurations. A wavelength of 0.8 nm was selected and the effective distances between the sample and the planar square multi detector (sample detector, S-D distance) were 2 and 8.0 m, respectively, with a 0.4 m detector offset for the 2 m S-D distance only. This allows for a momentum transfer range of 0.05 – 4 $nm^{-1}$. The instrument resolution is given by a 10 % wavelength spread and an uncertainty in angle $\Delta\theta = 2.1 \times 10^{-3}$ and $3.1 \times 10^{-3}$ for the 8.0 and 2 m S-D distance, respectively. The uncertainty in angle comprises contributions from the collimation, sample aperture, and detector cell size. The counting times were approximately 1 h/sample. The data from the remaining sample sets were collected with the PAXY diffractometer. A wavelength of 0.8 nm was selected and the S-D distances were 1.0 and 5.0 m, respectively. This allows for a momentum transfer range of 0.1 – 3 $nm^{-1}$. Here, the counting times per sample or solvent were approximately 4 and 7 hours for the 1.0 and 5.0 m S-D distance, respectively. Data correction allowed for sample transmission and detector efficiency. The efficiency of the detector was taken into account with the scattering of $H_2O$. Absolute intensities were obtained by reference to the attenuated direct beam and the scattering of the pure solvent with the same $H_2O/D_2O$ composition was subtracted. Finally, the data were corrected for a small solute incoherent contribution. To facilitate quantitative comparison of data collected with the D22 and PAXY instruments, we have done some duplicate measurements. Apart from a small difference in absolute normalization within 10%, the intensities obtained with the 2 different instruments are in perfect agreement. All data were corrected accordingly.

## C. X-ray scattering

For small angle x-ray scattering (SAXS), a set of solutions was prepared with 100% corona charge without added salt. To enhance the scattering contribution from the counterions, the copolymer was neutralized with CsOH. A range in concentration from 5 to 50 g of copolymer/l was obtained by concentrating a stock solution by means of evaporation in a vacuum oven in nitrogen atmosphere at reduced pressure in the presence of $P_2O_5$. For scattering, a sample droplet was deposited between mica sheaths in a sample holder and placed on a translation stage.

The synchrotron SAXS experiments were done at the BM26 "DUBBLE" beam line of the European synchrotron radiation facility (ESRF, Grenoble). The X-ray beam had a photon energy of 14.7 keV (wavelength $\lambda$ = 0.084 nm), bandwidth $\Delta\lambda/\lambda = 2\times 10^{-4}$ and a beam size of 100×100 $\mu m^2$ at the sample. Diffraction was detected at 8 m distance from the sample by a 13×13 $cm^2$ 2-dimensional gas-filled detector, which allows a momentum transfer range 0.06 – 0.9 $nm^{-1}$.

## D. Rheology

Flow curves for the 50% charged micelles were measured with a Contraves Low Shear 40 rheometer using a Couette geometry cell with inner and outer radii of 3.00 and 3.25 mm, respectively. The shear rate was varied between $5\times10^{-4}$ and 100 $s^{-1}$. The temperature of the cell was controlled at 298 K. It was checked that the viscosity was measured under steady-state conditions by monitoring the viscosity (at a constant shear rate) versus time. The visco-elastic moduli $G'(\omega)$ and $G''(\omega)$ were measured with a Bohlin VOR rheometer and a cone-plate geometry of diameter 60 mm and angle 1º. Data acquisition started when steady state was reached, as indicated by short test measurements of $G'$ and $G''$ at 1 Hz. Steady state was typically reached within 1 minute. Frequency sweeps were done between 0.001 and 10 Hz in the linear response regime. Prior to all measurements, the samples were pre-sheared for 5 minutes at a shear rate of 80 s$^{-1}$.

## V. RESULTS AND DISCUSSION

### A. Core and corona structure

With 4 experimental intensities pertaining to 4 different solvent compositions and 3 unknown partial structure factors, the SANS data are overdetermined and the structure factors can be obtained by orthogonal factorization in a least squares sense (*i.e.*, a 3-parameter fit to 4 data points for every value of $q$). However, the statistical accuracy of the derived partial structure factors can be improved if the structure factors are factorized into terms involving the radial core and/or corona profiles and a term describing the correlation of the center of mass of the micelles. As shown by Eq. (6), the intensities can than be expressed in terms of 2 unknown factors $u_i(q)$ rather than 3 partial structure factors $S_{ij}(q)$ (*i, j* = PS, PA). With a non-linear least-squares procedure, the 2 factors $u_i(q)$ were fitted to the data and the core and corona partial structure factors were reconstructed according to $S_{PS}(q) = u_{PS}^2(q)$ and $S_{PA}(q) = u_{PA}^2(q)$, respectively. Notice that with Eq. (4), the PS-PA cross structure factor does not carry additional information. In the low $q$ range, the standard deviation of the fit diverges and the intensities do not comply with solvent composition independent structure factors (not shown). This shows that the samples differ in secondary aggregation with concomitant long-range inhomogeneity in density, despite the fact that they have been prepared in the same way (but in various H$_2$O/D$_2$O solvent ratios). It was checked that for $q$ exceeding 0.07 nm$^{-1}$, the standard deviation has leveled off and the results are in perfect agreement with the model-free 3-parameter fit (but with improved statistical accuracy). As illustrative examples, the results pertaining to the 100 and 10% charged micelles without added salt, as well as fully charged micelles in 1 M KBr are displayed in Figs. 1-3, respectively.

The copolymer concentration covers the range from the diluted to the concentrated regime where the coronal layers have to shrink and/or interpenetrate in order to accommodate the micelles in the increasingly crowded volume. At the lowest micelle concentration and/or with excess salt, inter-micelle interference is insignificant and the core and corona structure factors can directly be compared with the relevant form factors. With increasing concentration and minimal screening conditions, a primary and higher order correlation peaks emerge. As is more clearly demonstrated by the SAXS results described below, the position of the primary peak scales with the copolymer concentration $C_{pol}$ according to $C_{pol}^{1/3}$. This scaling behavior is characteristic for micelles with fixed aggregation number and isotropic symmetry in the local environment. There are no major changes in the high $q$ behavior of the corona structure factor with increasing packing fraction, irrespective charge and ionic strength. This shows that the corona chain statistics is rather insensitive to inter-micelle interaction. For the 50 and 100% charged, salt-free micelles, the chains remain almost fully stretched and $\alpha = 2$. For the 10% charged micelles, the model calculations were done with $\alpha = 8/3$ in accordance with charge annealing towards the outer coronal region due to the recombination-dissociation balance of the weak polyacid. The lines in Figs. 1 and 2 represent the model calculations with form factor parameters $D_{core} = 9$ nm and $D_{mic}$ collected in Table III (the parameters pertaining to the fit of the center of mass structure factor, $D_{hs}$ and $\rho$, are discussed below). In the presence of excess salt, inter-micelle interference is largely suppressed and the corona structure factors are compared with the form factor calculated with $\alpha = 4/3$ (100 and 50% charge) or $8/3$ (10% charge). The results pertaining to the fully charged micelles in 1 M KBr are displayed in Fig. 3. All fitted micelle diameters are also collected in Table III.

With added salt and/or at low degree of ionization, the coronal layers are less extended. The ionic strength and charge dependencies of the micelle diameter agree with our previous results obtained for more diluted samples.[11,12,13] With increasing packing fraction, the diameter of the micelles, as obtained from the form factor analysis, decreases. However, the extent to which the coronal layers shrink is modest and similar under all circumstances. The gradual decrease in size is due to the restricted free volume, increased counterion adsorption, and/or Donnan salt partitioning between the coronal layer and the supporting medium.[6,7,8]

From the absolute normalization of the structure factors in the long wavelength ($q \to 0$), an aggregation number $N_{ag}$ around 100 is derived, irrespective of charge, copolymer concentration, and ionic strength.

## B. Counterion structure

The SAXS intensities of the fully neutralized PS-*b*-CsPA micelles divided by copolymer concentration are displayed in Fig. 4. For ease of comparison with the SANS data, the weight concentration refers to the copolymer only (*i.e.*, the concentrations are calculated without taking into account the relatively heavy Cs$^+$ counterion). Notice that the SAXS intensities are, to a good approximation, proportional to the counterion structure factor, because the scattering is dominated by the heavy Cs$^+$ ions. As in the case of the core and corona structure factor, the SAXS data show a primary inter-micelle correlation peak. The higher order correlation peaks are less prominent, due to the steep decrease of the counterion structure factor with increasing values of momentum transfer. As seen in the inset of Fig. 4, the position of the primary peak scales with copolymer concentration according to $C_{pol}^{1/3}$. This result confirms the isotropic local structure and fixed aggregation number of the micelles, which has also been observed with SANS. At high values of momentum transfer, the SAXS intensities are seen to increase with increasing concentration. This effect might be due to counterion fluctuations; a fit to a Gaussian background contribution yields a correlation length, which decreases from 1.2 to 0.7 nm with increasing copolymer concentration from 4.5 to 48 g/l, respectively.

As shown by previous SANS work of more diluted samples with isotopically labeled tetramethylammonium counterions, the distribution of the counterions along the micelle radius equals the one for the PA monomers.[13] Furthermore, from a quantitative comparison of the counterion and corona structure factors, it was concluded that, within a 10% error margin, all counterions are trapped in the coronal layer. Accordingly, it is interesting to compare the SAXS data for the PS-*b*-CsPA micelles with the relevant combination of the core PS and corona PA scattering contributions. The latter combination has been reconstituted with the SANS data from PS-*b*-NaPA solutions according to $(0.98\,u_{PA} + 0.02\,u_{PS})^2$ and divided by an arbitrary normalization factor. The 2% contribution from PS was optimized in a least-squares sense and accounts for the scattering from the core. As shown in Fig. 5, there is perfect agreement in both the position of the correlation peak and the variation of the structure factor with momentum transfer. Accordingly, the counterion distribution in the coronal layer is very close to the one of the corona forming polymer segments. This result agrees with our previous SANS work with isotopically labeled counterions as well as Monte Carlo simulation of urchin-like polyelectrolyte copolymer micelles.[13,14] It is also clear that the counterions remain strongly correlated with the coronal chains with increasing packing fraction up to and including the regime where the coronal layers interpenetrate. Notice that the present experiments do not allow an assessment of the extent to which the counterions are 2D localized around the stretched arms or 3D condensed inside the coronal layer. Due to the neutralization of the coronal layer by trapping of the counterions, the electrostatic contribution to the inter-micelle interaction potential is expected to play a minor role only.[17,18]

**C. Inter-micelle structure**

Inter-micelle interference is more clearly demonstrated in Fig. 6, where the core structure factor has been divided by the core form factor (10% and full charge and no added salt). Although the center of mass structure factor could also be derived from the corona structure factor, we have chosen to use the core structure factor because the core form factor shows a smooth and moderate variation in the relevant *q*-range (even so, consistency with the corona structure factor is illustrated in Figs. 1 and 2). The intensity of the correlation peaks first increases and eventually levels off with increasing packing fraction, which shows the progressive and saturating ordering of the micelles. Notice that for the present volume fractions the position of the primary peak is mainly determined by density, whereas the respective positions of the higher order correlation peaks are most sensitive to the value of the hard sphere diameter. The lines in Fig. 6 represent the hard sphere solution structure factor convoluted with the instrument resolution with fitted micelle densities and hard sphere diameters in Table III. The hard sphere model is capable of predicting the positions of the primary and higher order peaks. Furthermore, the ratio of the fitted micelle densities and known copolymer concentrations provides an alternative way to obtain the aggregation number. The aggregation numbers are also collected in Table III. The average value $N_{ag}$ = 98±10 is in perfect agreement with the one obtained from the normalization of the structure factors. The fixed functionality is evident from $C_{pol}^{1/3}$ scaling of the position of the correlation peak, the constant core diameter, and the constant aggregation number derived from both the absolute intensities and fitted micelle densities.

Clear deviations between the experimental data and the hard sphere prediction are observed in the low $q$-range. Furthermore, the model underestimates the intensity of the second order peak with respect to the primary one. We have checked that a repulsive, screened Coulomb potential does not improve the fit, nor does it significantly influence the peak positions for reasonable values of the micelle charge. The minor importance of the electrostatic interaction between the micelles and the relatively small net micelle charge due to the trapping of the counterions in the coronal layer are demonstrated by similar center of mass structure factors for 10, 50 and 100% charged micelles. The failure in predicting the relative amplitude of the higher order peak is probably related to the softness of the micelles; similar behavior has been reported for interpenetrating neutral polymer stars.[17,18] The deviations observed in the low $q$-range might be due to long-range inhomogeneity in density, the formation of aggregates, and/or stickiness between the micelles. We have checked that the sticky hard sphere model does not improve the fit in the low momentum transfer range, but it gives a better description of the depth of the first minimum after the primary peak (result not shown).[23] However, the derived distance of closest micelle approach is the same as the hard sphere diameter and we have further refrained from interpreting our data with this more elaborate model.

The outer micelle $D_{mic}$ and hard sphere $D_{hs}$ diameters as obtained from the form and solution structure factor analysis, respectively, are displayed in Fig. 7. $D_{mic}$ and $D_{hs}$ are estimated within a 3 and 2% accuracy margin, respectively, which is about the size of the symbols. The hard sphere diameters were derived for salt-free micelles only, because in the presence of excess salt inter-micelle interference is effectively suppressed. For the less concentrated, 15 and 17 g/l solutions, the hard sphere diameters equal the micelle diameters derived from the form factor analysis. This supports the applicability of the hard sphere interaction model in order to extract the effective hard sphere diameters. At higher packing fractions and for the 50 and 100% charged micelles in particular, the effective hard sphere diameters are significantly smaller than the outer micelle diameters. We take the difference as a measure of the extent to which the corona layers interpenetrate. Accordingly, the 100 and 50% charged, salt-free micelles interpenetrate around 17 g/l; for the smaller 10% charged micelles this happens at a higher concentration, say 25 g/l.

Fig. 7 also displays the average distance between the micelles $\rho^{-1/3}$. Once the micelles interpenetrate, the effective hard sphere diameter equals $\rho^{-1/3}$. Based on the optimized densities and hard sphere diameters, effective micelle volume fractions are calculated (Table III). For interpenetrating micelles, the effective volume fraction is found to be constant within experimental accuracy and takes the value $0.53\pm0.02$. Notice that, although this volume fraction corresponds with closely packed, simple cubic order, the center of mass structure factor remains liquid-like and no long-range order in the SAXS and SANS diffraction patterns is observed. Interdigitation thus occurs when the volume fraction exceeds the critical value 0.53. For higher copolymer concentration, this value is effectively preserved by interpenetration of the coronal layers.

## D. Visco-elastic behavior

The interpenetration of the coronal layers has a profound influence on the visco-elastic properties. All samples are fluid and flow when the test tubes are inverted. We have measured the viscosity of the solutions with 50% charged micelles. In excess salt (1 M KBr), the viscosity is in the range 1-2 mPa s, which is on the order of the viscosity of the solvent (data not shown). The viscosity versus shear rate of the samples without added salt is displayed in Fig. 8. A Newtonian plateau is observed, which increases in value by 3 orders of magnitude when the concentration is increased so that coronal layers interpenetrate. Notice that the salt-free sample with the lowest micelle concentration has already a 10 fold higher viscosity than the ones with excess salt. For the more concentrated samples, the onset of shear thinning at high shear rates is also observed.

We have also measured the dynamic moduli of the solutions with 50% charged micelles without added salt. Data are shown in Fig. 9. For the lowest concentration, the dynamic moduli show viscous liquid behavior with $G'(\omega) \sim \omega^2$ and $G''(\omega) \sim \omega^1$. For interpenetrating micelles and in the lower frequency range in particular, $G'(\omega)$ and $G''(\omega)$ show approximately parallel scaling laws as a function of frequency with scaling exponents around unity. Such behavior has been observed for a wide variety of polymer gels and similar micellar solutions of polyelectrolytes with adhesive corona chains.[2,29] Although we do not observe the transition to an elastic solid ($G' > G''$ and almost independent on frequency), an intuitive explanation of the parallel frequency scaling behavior of $G'$ and $G''$ is the formation of an interconnected network of micelles by the interpenetration of the coronal layers as shown by the scattering experiments.

## VI. CONCLUSIONS

With increasing packing fraction, the micelles shrink, irrespective corona charge and ionic strength of the supporting medium. The modest decrease in size with increasing concentration is due to the restricted free volume, increased counterion adsorption, and/or Donnan salt partitioning between the coronal layer and the supporting medium (the functionality is fixed due to the glassy core). The corona chain statistics is insensitive to inter-micelle interaction. For the 50 and 100% charged, salt-free micelles, the chains remain almost fully stretched. In the presence of excess salt and through the whole range of concentrations, the radial decay of the monomer density could be described by the same exponent as in neutral star-branched polymers (4/3). For the 10% charged micelles, the structure factors comply with annealing of the corona charge towards the outer region due to the recombination-dissociation balance of the weak polyacid. Irrespective the packing fraction, the counterions are strongly correlated with the corona forming segments with the same radial density profile. The present experiments do not allow an assessment of the extent to which the counterions are 2D localized around the stretched arms or 3D condensed inside in the coronal layer. However, almost all counterions are confined in either way, which results in a small net micelle charge.

Due to the relatively weak electrostatic interaction, inter-micelle correlation is insensitive to the corona charge fraction and can be satisfactorily described by a hard sphere model. From a comparison of the outer micelle and hard sphere diameters, as obtained from the form and solution factor analysis, respectively, it is concluded that the coronas of 50 and 100% charged, salt-free micelles interdigitate once the concentration exceeds 17 g/l. Based on the fitted hard sphere diameter, this concentration corresponds with an effective micelle volume fraction 0.53±0.02. At higher packing fractions, this critical value is effectively preserved by interpenetration of the coronal layers. For the smaller 10% charged micelles, interpenetration is observed at higher weight concentrations. Interpenetration of the polyelectrolyte brushes controls fluid rheology. As an example, the viscosity of the salt-free sample with 50% corona charge increases in value by 3 orders of magnitude, when the concentration is increased so that coronal layers interpenetrate. Furthermore, the parallel frequency scaling behavior of the dynamic moduli suggests the formation of an interconnected, physical gel. In the presence of excess salt and in the present concentration range, the coronal layers are less extended and they do not interpenetrate. Accordingly, the viscosity of the latter samples is in the range of the viscosity of the solvent.

## ACKNOWLEDGMENTS

We acknowledge the Institute Laue-Langevin, Laboratoire Léon Brillouin, and the European Synchrotron Radiation Facility in providing the neutron and X-ray research facilities, respectively. We thank the rheology group of the department of applied physics, Twente University for use of their rheology equipment. This research has been supported by the Netherlands Organization for Scientific Research and the European Community's access to large-scale facilities HPRI program (HPRI-CT-2002-00170).

TABLE I. Partial molar volumes and scattering lengths. The PA partial molar volumes were taken from Ref. [30]. $X$ denotes the D$_2$O mole fraction (effect of exchangeable hydrogen). The polymer data refer to the monomeric unit.

| Solute | $\bar{v}_i$ (cm$^3$/mole) | $b_i$ (10$^{-12}$ cm) |
|---|---|---|
| PAA | 48 | 1.66+1.04 $X$ |
| NaPA | 34 | 2.40 |
| KPA | 40 | 2.40 |
| PS | 99 | 2.33 |
| H$_2$O | 18 | -0.168 |
| D$_2$O | 18 | 1.915 |

TABLE II. Scattering length contrast in 10$^{-12}$ cm

| Solvent | $\bar{b}_{PS}$ | $\bar{b}_{PAA}$ | $\bar{b}_{NaPA}$ | $\bar{b}_{KPA}$ |
|---|---|---|---|---|
| H$_2$O | 3.2 | 2.1 | 2.7 | 2.8 |
| 29 % D$_2$O | 0.0 | 0.8 | 1.6 | 1.4 |
| 50 % D$_2$O ($DN$ = 0.1) | -2.4 | -0.1 | 0.7 | 0.5 |
| 70 % D$_2$O ($DN$ = 1) | -4.7 | -1.0 | 0.0 | -0.5 |
| 99 % D$_2$O | -8.0 | -2.3 | -1.2 | -1.8 |

TABLE III. Copolymer concentration $C_{pol}$, degree of neutralization $DN$, Salt concentration $C_s$, corona density scaling exponent $\alpha$, outer micelle diameter $D_{mic}$, effective hard sphere diameter $D_{hs}$, effective volume fraction $\eta_{eff}$, and aggregation number $N_{ag}$. The latter 3 entries can be estimated for the salt-free samples only (see text). The error margins in $D_{mic}$ and $D_{hs}$ are 3 and 2 %, respectively.

| $C_{pol}$ (g/l) | $DN$ | $C_s$ (M) | $\alpha$ | $D_{mic}$ (nm) | $D_{hs}$ (nm) | $\eta_{eff}$ | $N_{ag}$ |
|---|---|---|---|---|---|---|---|
| 4.4 | 1 | 0 | 2 | 48 | - | - | - |
| 17 | 1 | 0 | 2 | 45 | 45 | 0.43 | 116 |
| 30 | 1 | 0 | 2 | 42 | 38 | 0.51 | 102 |
| 44 | 1 | 0 | 2 | 40 | 34 | 0.53 | 102 |
| 4.5 | 1 | 1 | 4/3 | 33 | - | - | - |
| 17 | 1 | 1 | 4/3 | 32 | - | - | - |
| 30 | 1 | 1 | 4/3 | 31 | - | - | - |
| 44 | 1 | 1 | 4/3 | 30 | - | - | - |
| 4.5 | 0.5 | 0 | 2 | 45 | - | - | - |
| 17 | 0.5 | 0 | 2 | 43 | 42 | 0.43 | 104 |
| 30 | 0.5 | 0 | 2 | 42 | 36 | 0.54 | 90 |
| 44 | 0.5 | 0 | 2 | 38 | 33 | 0.55 | 99 |
| 4.5 | 0.5 | 1 | 4/3 | 30 | - | - | - |
| 17 | 0.5 | 1 | 4/3 | 27 | - | - | - |
| 30 | 0.5 | 1 | 4/3 | 25 | - | - | - |
| 43 | 0.5 | 1 | 4/3 | 22 | - | - | - |
| 4.5 | 0.1 | 0 | 8/3 | 42 | - | - | - |
| 15 | 0.1 | 0 | 8/3 | 41 | 41 | 0.50 | 82 |
| 25 | 0.1 | 0 | 8/3 | 38 | 37 | 0.51 | 88 |
| 44 | 0.1 | 0 | 8/3 | 34 | 33 | 0.55 | 101 |
| 4.4 | 0.1 | 0.04 | 8/3 | 33 | - | - | - |
| 15 | 0.1 | 0.04 | 8/3 | 31 | - | - | - |
| 25 | 0.1 | 0.04 | 8/3 | 28 | - | - | - |
| 44 | 0.1 | 0.04 | 8/3 | 27 | - | - | - |

FIG. 1. Core PS (a) and corona PA (b) structure factor versus momentum transfer for fully charged PS-*b*-PA micelles without added salt. The copolymer concentration is 44 (△), 30 (◇), 17 (□), and 4.4 (○) g/l from top to bottom. The data are shifted along the y-axis with an incremental multiplication factor. The curves represent the model calculations with core diameter 9 nm, and other parameters listed in Table III.

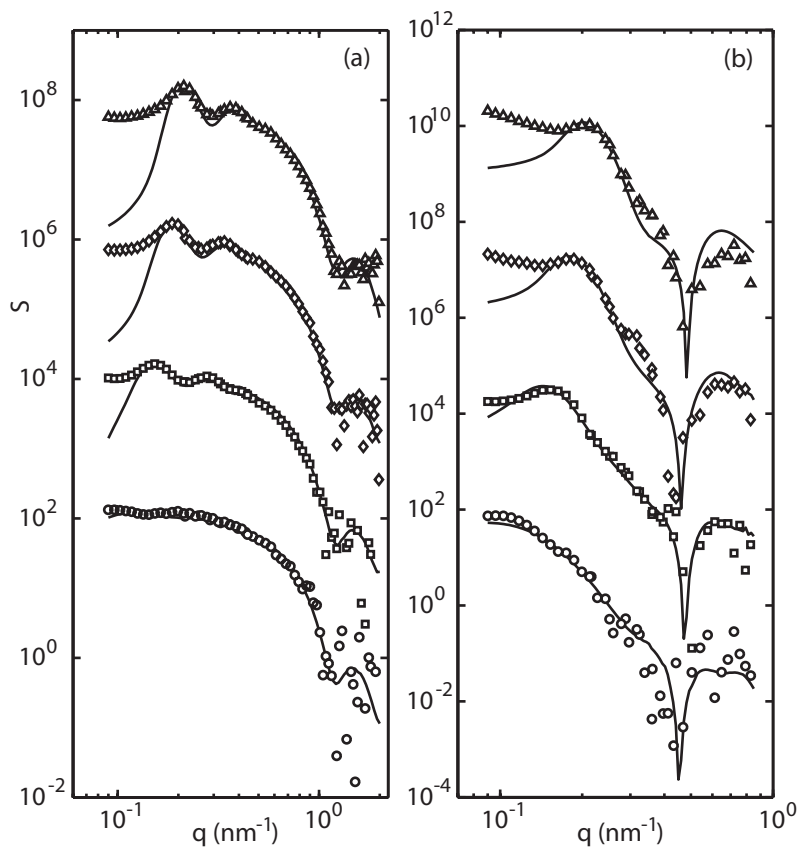

*Polyelectrolyte copolymer micelles*

FIG. 2. As in Fig. 1, but for 10% charged micelles without added salt. The copolymer concentration is 44 (△), 25 (◇), 15 (□), and 4.5 (○) g/l from top to bottom.

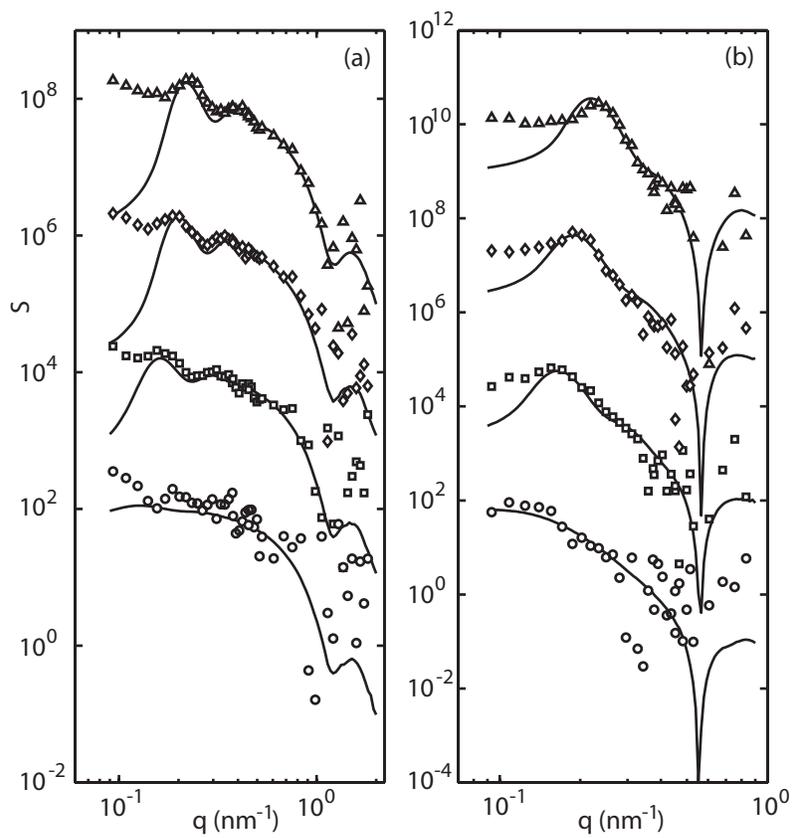

FIG. 3. As in Fig. 1, but for fully charged micelles in 1 M KBr. The copolymer concentration is 44 (△), 30 (◇), 17 (□), and 4.5 (○) g/l from top to bottom.

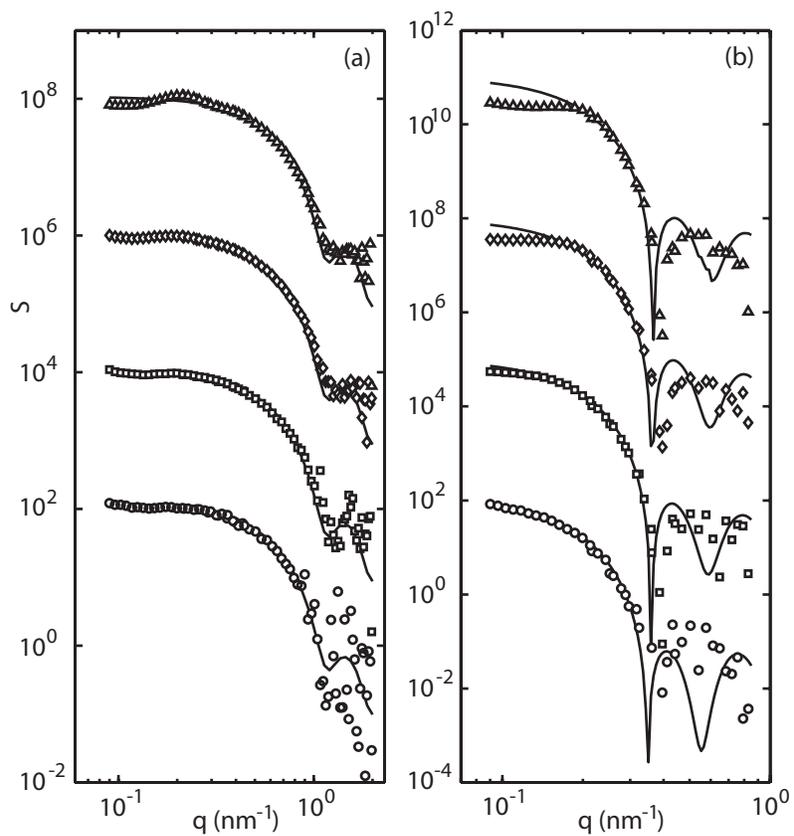

*Polyelectrolyte copolymer micelles*

FIG. 4. SAXS intensity divided by copolymer concentration versus momentum transfer for fully charged PS-*b*-CsPA micelles without added salt. The inset shows the peak position $q_m$ versus copolymer concentration $C_{pol}$ in the double logarithmic representation. The line represents $q_m \sim C_{pol}^{1/3}$ scaling.

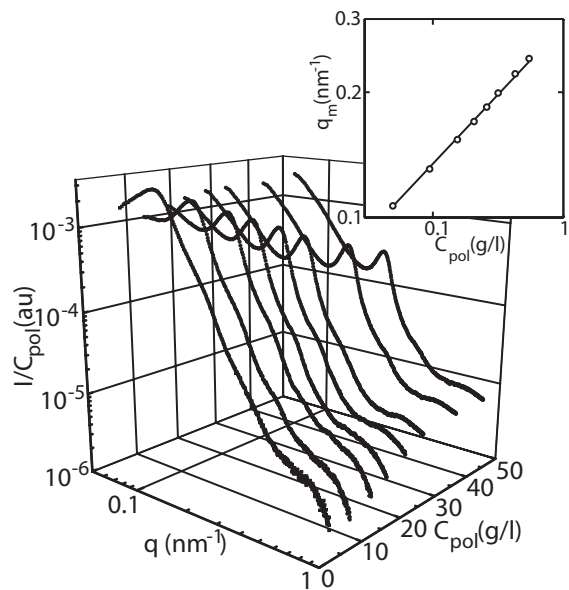

FIG. 5. Comparison of SAXS (solid curves) intensities with the reconstituted SANS intensities (symbols). The data are shifted along the y-axis with an incremental multiplication factor. The copolymer concentration is 44 (for SAXS 48, △), 30 (◇), 17 (□), and 4.5 (○) g/l from top to bottom. For the most densely concentrated set, notice the slight difference in position of the correlation peak due to a small difference in micelle concentrations.

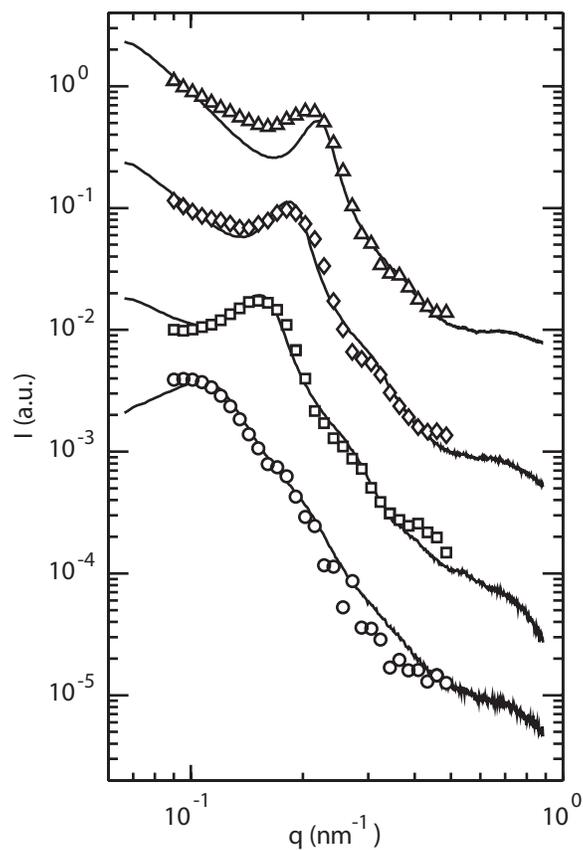

*Polyelectrolyte copolymer micelles*

FIG. 6. Center of mass solution structure factor for 10% (a) and fully (b) charged PS-*b*-PA micelles versus momentum transfer. The concentrations are as in Figs. 1 and 2. The data are shifted along the y-axis with an increment of 1.5 units. The curves represent the hard sphere solution structure factor with parameters listed in Table III. Notice that for the 10% charged micelles the effective volume fractions are slightly larger (Table III), which results in a somewhat more intense correlation peak in comparison with the one for the fully charged micelles.

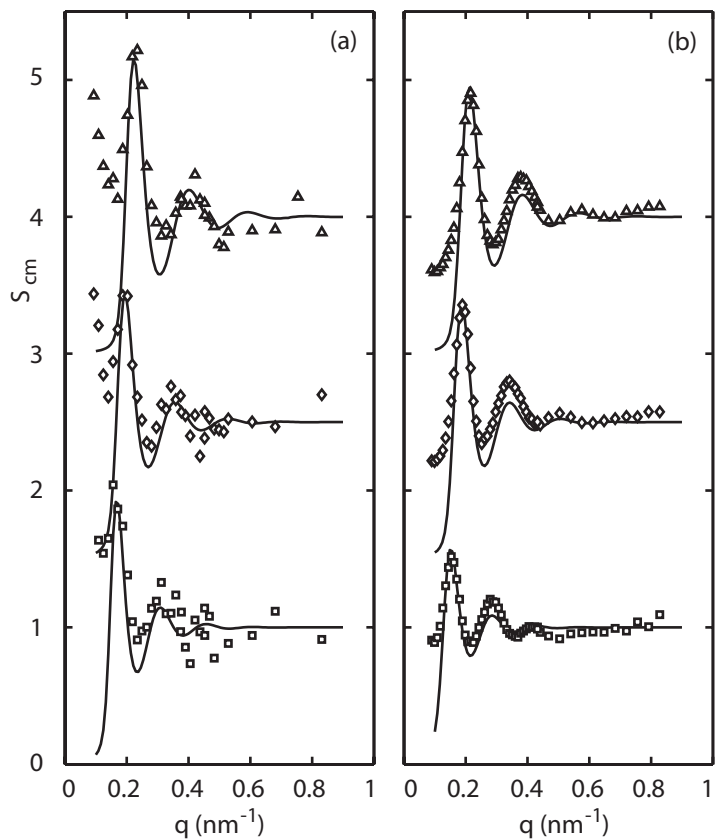

FIG. 7. Concentration dependence of the diameter of 100 (a), 50 (b), and 10% (c) charged PS-*b*-PA micelles: (●), $D_{mic}$ without added salt; (○), $D_{mic}$ in 1.0 M (100 and 50%) or 0.04 M (10% charge) KBr. The hard-sphere diameter of salt-free micelles $D_{hs}$ is denoted by (□). The lines represent $\rho^{-1/3}$, *i.e.* the average inter-micelle distance.

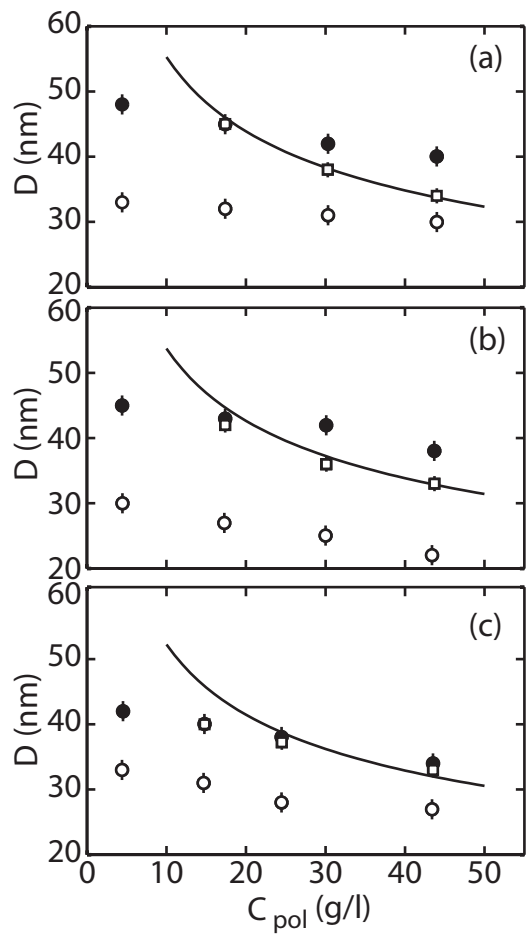

FIG. 8. Viscosity versus the shear rate for 50% charged PS-*b*-PA micelles without added salt. The copolymer concentration is 44 (△), 30 (◇), 17 (□), and 4.5 (○) g/l from top to bottom.

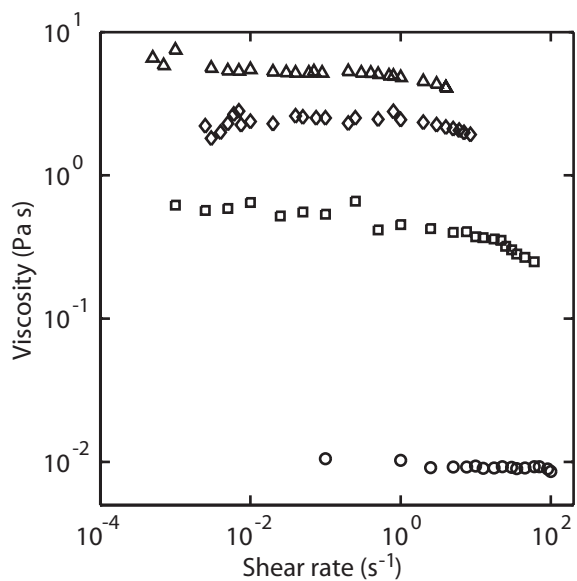

*Polyelectrolyte copolymer micelles*

FIG. 9. Frequency dependence of storage $G'$ (open symbols) and loss $G''$ (filled symbols) modulus for 50% charged PS-*b*-PA micelles without added salt. The copolymer concentration is 44 (△), 30 (◇), 17 (□), and 4.5 (○) g/l from top to bottom. The lines indicate $\omega^1$ and $\omega^2$ scaling.

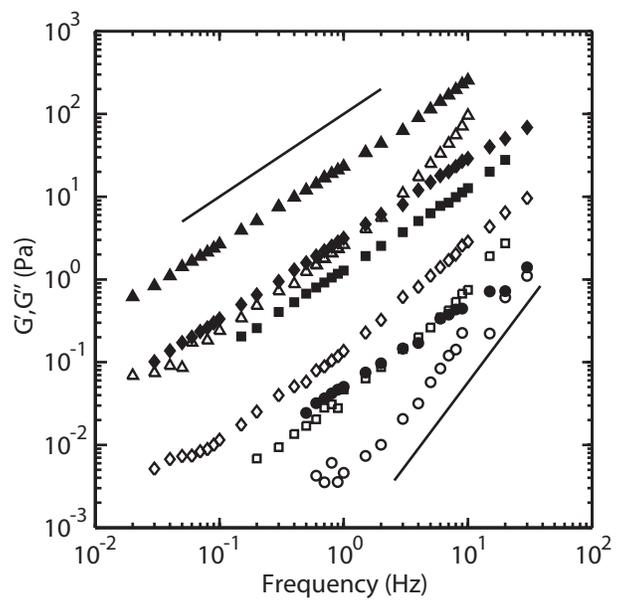

---

[19] Note that this definition differs from the situation for homopolymers, where the monomer is usually considered the elementary scattering unit.

[20] J.K. Percus and G.J. Yevick, Phys. Rev. **110**, 1 (1958).

[21] C.N. Likos, H. Lowen, M. Watzlawek, B. Abbas, O. Jucknischke, J. Allgaier, and D. Richter, Phys. Rev. Letters **80**, 4450 (1998).

[22] J.B. Hayter and J. Penfold, Mol. Phys. **42**, 109 (1991).

[23] R.J. Baxter, J. Chem. Phys. **49**, 2770 (1968).

[24] J.S. Pedersen and M.C. Gerstenberg, Macromolecules **29**, 1363 (1996).

[25] S. Förster and C. Burger, Macromolecules **31**, 879 (1998).

[26] L. Auvray, C. R. Acad. Sc. Paris **302**, 859 (1986).

[27] L. Auvray and P. G. de Gennes, Europhys. Lett. **2**, 647 (1986).

[28] M. Daoud and J.P. Cotton, J. Phys. (Paris) **43**, 531 (1982).

[29] H.H. Winter and M. Mours, Adv. Polym. Sci. **134**, 165 (1997).

[30] K. Hiraoka and T. Yokoyama, J. Polym. Sci.: Polym. Phys. **24**, 769 (1986).